\documentclass[fleqn,usenatbib]{mnras}

\usepackage{hyperref}
\usepackage{newtxtext,newtxmath}
\usepackage[T1]{fontenc}
\usepackage{ae,aecompl}

\usepackage{graphicx}	% Including figure files
\usepackage{amsmath}	% Advanced maths commands
\usepackage{amssymb}	% Extra maths symbols
\usepackage{subfig}
\usepackage[wby]{callouts}
\usepackage{pdflscape} % for 'landscape' environment
\usepackage{numprint} %for rounding
\usepackage{longtable} % for 'longtable' environment

\graphicspath{ {images/} }

\def\Fig{\mbox{Figure~}}
\def\Figs{\mbox{Figures~}}
\def\Tab{\mbox{Table~}}
\def\Sec{\mbox{Section~}}

\def\Mauto{\mbox{{\tt MAG\_AUTO}}}
\def\Mautor{\mbox{{\tt MAG\_AUTO\_r}}}
\def\maggaap{\mbox{{\tt MAG\_{GAaP}}}}
\def\maggaapr{\mbox{{\tt MAG\_{GAaP}\_r}}}

\def\Msun{\mbox{${\rm M}_\odot$}}

\def\Zsun{\mbox{${\rm Z}_{\odot}$}}

\def\lsim{\mathrel{\rlap{\lower3.5pt\hbox{\hskip0.5pt$\sim$}}
    \raise0.5pt\hbox{$<$}}}% less than or approx. symbol
\def\gsim{~\rlap{$>$}{\lower 1.0ex\hbox{$\sim$}}}

\newcommand{\new}[1]{{}}
\newcommand{\newcre}[1]{#1}
\title[LinKS]{LinKS: Discovering galaxy-scale strong lenses in the Kilo-Degree Survey using Convolutional Neural Networks}

\author[C.~E.~Petrillo et al.]{C.~E.~Petrillo$^{1}$\thanks{E-mail: petrillo@astro.rug.nl},  C.~Tortora$^{1,2}$, G.~Vernardos$^{1}$,   L.~V.~E.~Koopmans$^{1}$, \and G.~Verdoes~Kleijn$^{1}$, M.~Bilicki$^{3,4}$, N.~R.~Napolitano$^{5}$, S.~Chatterjee$^{1}$, G.~Covone$^{6}$, \and A. Dvornik$^{3}$, T. Erben$^{7}$, F.~Getman$^{5}$, B. Giblin$^{8}$,   C.~Heymans$^{8}$, J.~T.~A.~de~Jong$^{1}$, \and K.~Kuijken$^{3}$, P.~Schneider$^{7}$, H.~Shan$^{9}$, C. Spiniello$^{5}$, A.~H.~Wright$^{7}$\vspace{0.2cm} \\ 
$^{1}$Kapteyn Astronomical Institute, University of Groningen, Postbus 800, 9700 AV, Groningen, The Netherlands\\
$^{2}$INAF -- Osservatorio Astrofisico di Arcetri, Largo Enrico Fermi 5, 50125, Firenze, Italy\\
$^{3}$Leiden Observatory, Leiden University, P.O.Box 9513, 2300RA Leiden, The Netherlands\\
$^{4}$Center for Theoretical Physics, Polish Academy of Sciences, al. Lotnik\'{o}w 32/46, 02-668, Warsaw, Poland\\
$^{5}$INAF -- Osservatorio Astronomico di Capodimonte, Salita Moiariello, 16, 80131 Napoli, Italy\\
$^{6}$Dipartimento di Scienze Fisiche, Universit\`a di Napoli Federico II, Compl. Univ. Monte S. Angelo, 80126 Napoli, Italy\\
$^{7}$Argelander-Institut für Astronomie, Auf dem H\"ugel 71, 53121 Bonn, Germany\\
$^{8}$Institute for Astronomy, University of Edinburgh, Royal Observatory, Blackford Hill, Edinburgh, EH9 3HJ, UK\\
$^{9}$Shanghai Astronomical Observatory (SHAO), Nandan Road 80, Shanghai 200030, China
}

\date{Accepted XXX. Received YYY; in original form ZZZ}

\pubyear{2018}

\begin{document}
\label{firstpage}
\pagerange{\pageref{firstpage}--\pageref{lastpage}}
\maketitle

\begin{abstract}
We present a new sample of galaxy-scale strong gravitational-lens candidates, selected from 904 square degrees of Data Release 4 of the Kilo-Degree Survey (KiDS), i.e., the ``Lenses in the Kilo-Degree Survey" (LinKS) sample. We apply two Convolutional Neural Networks (ConvNets) to $\sim88\,000$ colour-magnitude selected luminous red galaxies yielding a \newcre{list of 3500} strong-lens candidates. This list is further down-selected via human inspection. The resulting LinKS sample is composed of 1983 rank-ordered targets classified as ``potential lens candidates" by at least one inspector.  Of these, a high-grade subsample of 89 targets is identified with potential strong lenses by all inspectors. Additionally, we present a collection of another 200 strong lens candidates discovered serendipitously from various previous ConvNet runs. A straightforward application of our procedure to future Euclid \newcre{or LSST} data can select a sample of $\sim3000$ lens candidates with less than 10 per cent expected false positives and requiring minimal human intervention.
\end{abstract}


%%%%%%%%%%%%%%%%%%%%%%%%%%%%%%%%%%%%%%%%%%%%%%%%%%

%%%%%%%%%%%%%%%%% BODY OF PAPER %%%%%%%%%%%%%%%%%%

\section{Introduction}

Strong gravitational lenses\footnote{Called strong lenses or simply lenses hereafter.} are composite systems where a massive foreground object (e.g., a galaxy or a cluster) creates multiple images of one or more higher-redshift sources (e.g., galaxies or quasars). Strong lenses are useful for a wide range of cosmological and astrophysical studies (\citealt{schneider1992gravitational}; \citealt{Schneider06_SAASFEE33};\citealt{Treu2010rev}). For example, they can provide cosmological constraints on the dark energy equation of state (\citealt{Collet2014de,Cao2015de}) and precision measurements of the Hubble constant \citep{Schechter1997,Suyu2013,bonvin2016}. 
The information obtained from strong lensing also allows us to study the mass distribution in the inner regions of galaxies: e.g., the fraction of dark matter in their central regions (\citealt{gavazzi2007, jiang2007, grillo2010, Cardone2010, tortora2010central, more2011, ruff2011, sonnenfeld2015sl2s}), the slope of their inner mass density profile \citep{Treu2002MNRAS, koopmans2006, Moore2008, barnabe2009, koopmans2009, Shu2015, cao2016limits,Li2018} and their dark-matter substructures \citep{More2009, Vegetti2012, Nierenberg2014, Hezaveh2016}. Besides studying dark matter,  strong lenses allow us to place constraints on the stellar Initial Mass Function (IMF) when combined with dynamical and stellar population synthesis analyses (\citealt{treu2010, ferreras2010, Spiniello:2011p8239, brewer2012swells,Barnabe2013,sonnenfeld2015sl2s,posacki2015stellar,Spiniello+15_IMF_vs_density,leier2016strong,Sonnenfeld2018imf,Vernardos2018}).
Finally, strong lenses can act as a ``Cosmic Telescope'', providing a magnified view of otherwise unresolved background sources (e.g., \citealt{impellizzeri2008,swinbank2009,richard2011,deane2013,treu2015grism,Mason2016,Salmon2017,Kelly2017}).

The above-listed studies have typically been carried out using samples of tens to maximally about a hundred massive lens galaxies  (${\rm M}_\star \gtrsim 10^{11}\Msun$), and are often limited to redshifts $z \lesssim 0.5$ \newcre{and/or are inhomogeneously selected}. Current results are therefore often limited by sample size or cosmic variance. Creating more substantial, homogeneously selected samples of gravitational lenses, which extend to lower-mass galaxies and higher redshifts, will reduce the effects of ``small-number statistics" and allow an improved study of lens galaxies as a function of galaxy properties and evolutionary state.
In particular, \cite{Vegetti2009} estimate that it is possible to compute sub-halo mass fractions of lens galaxies to a level of $\lesssim 0.1$ per cent with only $\sim50$ lens systems.
With the same number of lenses, it is possible to reach a per cent level precision in estimating their mass density slopes \citep{Barnab2011}. Therefore a much larger number of galaxy-scale lenses can improve the outcome from these analyses and enable one to conduct a proper statistical comparison with the results obtained from lens simulations (e.g., \citealt{Xu2016, Li2016, Mukherjee2018}). Moreover, the precision of the value of $H_0$ can be improved to the level of a few per cent when studying a sample of about 40 strong lenses \newcre{with measured time delays} \citep{Jee2016, Shajib2018}. Collecting large samples of strong lenses, furthermore, giving us better access to the high-redshift universe and increases the probability of discovering double Einstein-ring \citep{Gavazzi2008} and other ``exotic" lenses  (e.g., \citealt{Tu2009, Cooray2011, Brammer2012, Tanaka2016}). 
Moreover, samples of homogeneously selected strong lenses are needed to characterize the selection function of a strong lens survey, allowing to map measurements carried out on strong lenses back to the general population of galaxies.
We refer the reader to the LSST Science Book \citep{abell2009} and the Euclid Strong Lensing white paper (Euclid Strong Lensing team, 2018, in prep) for a more detailed discussion of future scientific applications of strong gravitational lenses. 

The largest homogeneously-selected sample of confirmed strong lenses is the Sloan Lens ACS Survey (SLACS; \citealt{Bolton2006, SLACS2008}), which yielded more than a hundred spectroscopically confirmed strong lenses with complete redshift information and high-resolution imaging follow-up (with e.g., the Hubble Space Telescope and Keck Observatory Adaptive Optics). In total, all lens surveys combined have produced up to a thousand highly-likely\footnote{Not all of these lenses have been spectroscopically confirmed though, but from their image geometry are extremely probable to be strong lenses.}  gravitational lens candidates (e.g., \citealt{Browne2003, Faure2008, Treu2011, Inada2012, Brownstein2012, More2012, Stark2013, Sonnenfeld2013, Gavazzi2014, more2016,Shu2016,Shu2017}).

Ongoing wide-field optical-IR surveys are expected to make the next giant step forward by yielding thousands of new lenses \citep{collet2015, Petrillo2017}. The first new lens candidates have already been discovered \citep{Petrillo2017, Hartley2017, Diehl2017, Sonnenfeld2018, Spiniello2018,Jacobs2018,Wong2018} in the Kilo-Degree Survey (KiDS; \citealt{deJong2013}), in the Hyper Suprime-Cam Subaru Strategic Program (HSC; \citealt{HSCmiyazaki2012hyper}), and in the Dark Energy Survey (DES; \citealt{DES}). Similarly large samples are expected from deep sub-mm observations by e.g., the Herschel telescope \citep{negrello2010}, the South Pole Telescope (SPT; \citealt{carlstrom2011}), and the Atacama Large Millimeter/sub-millimeter Array (ALMA)\footnote{\href{http://www.almaobservatory.org/}{\tt http://www.almaobservatory.org/}}. These telescopes have already uncovered hundreds of new lens candidates \citep{vieira2013dusty, negrello2017herschel}.
Within the next decade, $\sim 10^5$ strong lenses are expected to be found in future surveys \citep{oguri2010gravitationally,pawase2012,collet2015,McKean2015} utilising, e.g., ESA's Euclid mission \citep{Laureijs:2011wi}, the Large Synoptic Survey Telescope \citep{abell2009} and the Square Kilometer Array\footnote{\href{https://www.skatelescope.org/}{\tt https://www.skatelescope.org/}}. 
In particular, these surveys will allow lower-mass and higher-redshift lenses to be found, thanks to their deeper and higher angular resolution observations. Moreover, it will become possible to follow up promising targets at an even higher angular resolution with ALMA and the European Extremely Large Telescope (E-ELT). A future SKA-VLBI facility could, in addition, investigate milli-arcsecond angular scales of the lensed images for the effects of dark-matter line-of-sight and sub-halos \citep{Spingola2018}, enabling one to study small deviations from the smooth mass model of the lens.

Strong gravitational lenses are scarce objects within the total population of galaxies. In current surveys, of the order of one strong lens exists per few hundred to a thousand galaxies. This number strongly depends \newcre{on galaxy mass and selection criteria, with the number of lenses peaking around ${\rm M}^*$-galaxies for source-selected samples and at larger masses when lenses are selected as luminous red galaxies (LRGs).} Their rarity makes it essential to develop robust lens-finder algorithms and deploy them in streamlined data-processing pipelines. This end-to-end automation will drastically reduce, and possibly prevent entirely, the need for future visual inspection of millions of potential lens candidates (e.g., \citealt{Lenzen2004, Horesh2005, Alard2006, Estrada2007, Seidel2007, Kubo2008, More2012, Maturi2014, Joseph2014, Gavazzi2014, Agnello2015, Brault2015, Chan2015, Stapelberg2019, Hartley2017, Petrillo2017, Petrillo2019, Jacobs2017, Sonnenfeld2018, Spiniello2018}).

In light of such an automation strategy,  we recently developed \citep{Petrillo2017}, and more recently improved upon \citep{Petrillo2019}, a new convolutional neural network (ConvNet) lens-finder algorithm. The objective in this paper is to report on how we use ConvNets in an automated lens-search pipeline, and report on the results of applying these networks to galaxies selected from $\sim900$ square degrees of KiDS Data Release 4. The core result that we present is an automatically selected sample of 3500 rank-ordered strong-lens candidates. From this ConvNet pre-selected sample, several subsamples of higher confidence candidates are distilled through human visual inspection. 

In \Sec\ref{SECdata}, we provide a brief introduction to KiDS, the imaging and catalogue data that are used in this paper. In \Sec\ref{SECfindlenses}, we explain how we select a subsample of intrinsically luminous (red) galaxies from the colour-magnitude diagram of KiDS galaxies, as well as the methodology used to identify gravitational lens candidates within that colour-magnitude selected subsample.  In \Sec\ref{SEClinks}, we present the gravitational lens candidates found from the most conservative sample selection. In \Sec\ref{SECfullsurvey}, we apply the networks to a wider selection of galaxies -- inherently limited only in their apparent brightness -- to examine the efficiency of the algorithm in extremely data-heavy regimes such as those expected from future astronomical surveys, such as with Euclid \newcre{and LSST}, which may also have restricted colour information. \newcre{In the same section we also present} a ``bonus sample" of inhomogeneously selected lens candidates that were identified serendipitously during various past experiments in the development of the final ConvNets. Lastly, in \Sec\ref{SECdiscussion}, we summarise our main conclusions. 

\section{Data from the Kilo-Degree Survey}\label{SECdata}

The Kilo-Degree Survey (KiDS; \citealt{deJong2013}) is an ESO public survey carried out with the OmegaCAM wide-field imager 
(\citealt{Kuijken11}) mounted on the VLT Survey Telescope (VST; \citealt{Capaccioli_Schipani11})  at the Paranal Observatory in Chile. The telescope, camera, and survey have been designed to obtain images with sub-arcsecond seeing and homogeneous image quality both across the full field of view and throughout the survey execution. In this way the survey yields a large and homogeneous galaxy sample. The size and homogeneity of this sample is required for the surveys primary science   drivers, which include placing strong constraints on both the distribution of matter across cosmic time and the cosmological parameters of the universe through weak-lensing measurements; the subtle distortions introduced in galaxy shapes by cosmic shear (e.g., \citealt{Hildebrandt2017}). At the same time, the combined power of the survey's superb image quality and wide area makes KiDS optimal for strong-lensing studies \citep{Napolitano+16_lensing,Petrillo2017,Spiniello2018}. OmegaCAM has a one square degree field of view, with pixels that have an angular scale of 0.21 arcseconds, and KiDS will survey  a total of $\sim$1350 square degrees in four optical bands ($u$, $g$, $r$ and $i$) by the end of observations in 2019.  
The best seeing observations are reserved for the $r$-band, with the survey exhibiting median point spread function (PSF) full-width at half-maximum (FWHM) values of 1.0, 0.8, 0.65 and 0.85 arcseconds in the $u-$, $g-$, $r-$, and $i-$bands respectively. The survey depths per-band, as determined by the $5-\sigma$ limiting magnitudes within a 2 arcsecond circular aperture, are 24.2, 25.1, 25.0, 23.7 in the $u-$, $g-$, $r-$, and $i-$bands respectively \citep{deJong+15_KiDS_paperI,deJong+17_KiDS_DR3}. 

In this paper, we make use of 904 tiles\footnote{The full fourth KiDS data release consists of 1006 tiles, but we have chosen to limit our analysis to the first 904 tiles that were processed by Astro-WISE.} that form a subset of the KiDS Data Release 4 (KiDS ESO-DR4, \newcre{Kuijken et al. 2018, in prep.}). \newcre{The analysis performed uses} imaging data, and derived products, produced within the Astro-WISE information system \citep{Valentijn2007, McFarland2013}. 
We make use of the single-band and multi-band catalogues of the KiDS-DR4.

\subsection{The ``full sample"}
 
The target extraction and their associated photometry have been obtained using \textsc{S-Extractor} (\citealt{Bertin_Arnouts96_SEx}). 
To optimise the initial lens searches, we pre-select a sample of luminous galaxies with reliable photometric data. We proceed in the following way: 

\begin{enumerate}

\item[(a)] \noindent We select sources with a \textsc{S-Extractor} $r$-band {\tt FLAGS} value $< 4$, thereby including only deblended sources and removing from the catalogue objects with incomplete or corrupted photometry, saturated pixels, or any other blending or extraction related problem.
 
\item[(b)] \noindent We further reject galaxies in areas compromised by, e.g, stellar diffraction spikes and reflection halos, by selecting sources with the flag {\tt IMA\_FLAGS} set to zero for all the four KiDS bands.

\item[(c)] \noindent We select sources with a Kron-like magnitude \Mauto\ in the $r$-band below 20th magnitude, in order to maximize the lensing cross-section (\citealt{schneider1992gravitational}).

\item[(d)] \noindent Finally, we select sources with flag {\tt 2DPHOT} equal to 1 (as derived by the star-galaxy separator software \textsc{2DPHOT} \citep{LaBarbera2008} in order to \newcre{select secure} galaxies. 

\end{enumerate}

\noindent To reduce the contamination by stars further, we select only objects with a FWHM in $r$-band greater than the 90 percentile range of the distribution of star-like objects within the same tile (those with {\tt 2DPHOT} equal to zero). We adopt this strategy to reach a suitable compromise between filtering out stars and not excising too many galaxies from the sample.
This selection procedure results in a sample of nearly one million (specifically 930\,651) targets which we will refer to as the ``\textit{full sample}'' in the remainder of the paper.

\subsection{The luminous red galaxy sample}\label{sec:LRG}

Luminous Red Galaxies (LRGs; \citealt{Eisenstein2001}) are massive galaxies which, as a result, are more likely to exhibit strong lensing features than other classes of galaxies (see \citealt{turner1984statistics, fukugita1992statistical, kochanek1996flat, chae2003cosmic, oguri2006image, moller2007strong}). We select LRGs from the \textit{full sample}, defined earlier, using the low-redshift ($z<0.4$) LRG colour-magnitude selection of \cite{Eisenstein2001}. We slightly adapt this selection to include fainter and bluer sources:%\footnote{After the analysis for this project began, a more sophisticated selection of LRG galaxies, for clustering studies, has been carried out by \cite{Vakili2018} for 440 survey tiles.}: 
\begin{equation}
\begin{split}
&|c_{\rm{perp}}| < 0.2\,,\\
&r<14+c_{\rm{par}}/0.3 \\
\text{where}\\
&c_{\rm{par}}=0.7(g-r)+1.2[(r-i)-0.18)]\,,\\
&c_{\rm{perp}}=(r-i)-(g-r)/4.0-0.18\,.
\end{split}
\end{equation}
The magnitudes are \textsc{S-Extractor} \Mauto.
In this section we chose to limit our analysis to the {\sc Astro-WISE} single-band object detection catalogues. We determine the {\it u,g,r,i} photometry for each object using the individual {\sc S-Extractor}  {\tt MAG\_AUTO} measurements.
As these measurements are made using slightly different centroids and the PSF varies significantly between bands, we do not expect this ``first-look" LRG selection methodology to be uniform.  As our aim is not to compile a complete sample of LRGs, however, we do not expect this decision to impact our conclusions.   We note that after the analysis for this project began, \cite{Vakili2018} presented a sophisticated methodology to select LRG galaxies for clustering studies in KiDS-DR3.  Future LinKS analyses will investigate adopting this LRG sample.
Our selection results on a sample of 88\,327 sources, which we refer as the ``\textit{LRG sample}" throughout the remainder of this paper. Note that our goal here is to select a reasonable number of massive (LRG) galaxies, without significant contamination by spiral galaxies, but that this sample need not strictly be purely LRGs. We find an average of 98 sources selected per tile with a standard deviation of $\sim43$. 
This standard deviation is high, but expected given the ``first-look" methodology that we have adopted to compile this sample, in addition to the high levels of cosmic variance expected for this highly biased galaxy sample.
%The standard deviation is considerably larger than the expected one based purely on Poisson noise and cosmic variance, this is most likely due to non-Gaussian cosmic variance. For example, \cite{Tegmark2006} find that clustering of LRGs suggests a non-linear bias of a factor $\sim3.6$ relative to the dark-matter mass power spectrum at $z=0$. Such a bias could explain the larger tile-to-tile variance than is expected from simple Poisson statistics. Also \cite{Fassnacht2006, Fassnacht2011} found that lens galaxies, such as LRGs, are generally highly clustered. \newcre{Nonetheless, we cannot exclude the possibility that some fraction of this large scatter may be due to \Mauto\ magnitudes being biased by PSF variations across the KiDS area.}

\section{Searching for Lenses}\label{SECfindlenses}

To find gravitational-lens candidates in KiDS imaging data, we use the ConvNets previously introduced by \cite{Petrillo2019}. These networks are significantly improved variants of the original ConvNet presented by \cite{Petrillo2017}. 
ConvNets \citep{fukushima1980neocognitron, lecun1998gradient} represent a state-of-the-art method of pattern recognition \citep{ILSVRC15}. The networks \textit{learn} how to classify a diverse set of images during the so-called training phase, whereby labelled images are provided to the ConvNet. \newcre{Its weight} parameters are changed to minimise a pre-defined loss function, which  expresses the difference between the labels of the images and the output values $p$ (one for each image) of the ConvNet. For a more detailed introduction to ConvNets for finding lenses we refer the interested reader to \cite{Petrillo2017}, and to more general reviews by \cite{schmidhuber2015deep}, \cite{lecun2015deep} and \cite{guo2016deep}.

To evaluate methods for identifying images of simulated gravitational lenses -- in preparation for the Euclid mission \citep{Metcalf2018} -- recently an international challenge was organised. The results of this challenge demonstrated that ConvNets, collectively with Support Vector Machines (SVMs), are among the most promising methods for finding lens candidates currently available.  As a proof of concept, ConvNets have been used to find new gravitational lens candidates by \cite{Petrillo2017} in the KiDS DR3 and by \cite{Jacobs2017} in the Canada-France-Hawaii Telescope Legacy Survey (CFHTLS) and in DES \citep{Jacobs2018}.  

In terms of methodology and target selection our analysis differs from the work done by \cite{Spiniello2018}, who have focused their search exclusively on lensed quasar candidates in KiDS, by visual inspecting targets preselected using optical/infrared colours. 
Lens candidates have also been found in the KiDS DR3 data by \cite{Hartley2017}, who trained a Gabor-SVM finder.

\subsection{Training the Convolutional Neural Networks}

We start by giving a brief synopsis of our ConvNets and the training procedure, as reported by \cite{Petrillo2019}. Building on our experience, we choose to deploy two different ConvNets. One focusses on utilising the best morphological information by taking the best-seeing, i.e.,  \textit{r}-band, images as input. \newcre{The other ConvNet exploits} colour information in addition to morphological information by taking 3-band RGB images as input. The RGB images are created with \textsc{HumVI}\footnote{\href{https://github.com/drphilmarshall/HumVI}{\tt https://github.com/drphilmarshall/HumVI}} \citep{Marshall2016} using the \textit{g}, \textit{r} and \textit{i} bands. In both cases, the KiDS images have a size of $101\times101$ pixels \newcre{(i.e. $20 \times 20$ arcseconds)} with the central pixel corresponding to the centre of the galaxy of interest. The ConvNets take these images and transform them into a single value, $p$, which can vary between 0 and 1. This value represents, to some degree \citep[see e.g.,][]{saerens2002}, the probability that the input image is a lens (see also Section 3.2). The input size of $20 \times 20$ arcseconds is chosen to be sufficiently large as to enclose most galaxy-scale lens systems, and sufficiently small as to both avoid contamination by unrelated field objects and allow for a ConvNet with a practical memory requirement\footnote{Larger images require a larger numbers of network weights and consequently more computer memory.}.

We use two classes of objects to train the ConvNets: (1) the \textit{lenses} labelled with a 1.0, and (2) the \textit{non-lenses}, labelled with a 0.0.  
\begin{itemize}
\item[(1)] For the \textit{lenses}, we use a set of $\sim6000$ KiDS LRGs on which we superimpose simulated lensed images. The simulated lensed images ($\sim10^6$ in number) are composed mostly of high-magnification rings, arcs and quads. The gravitational-lens mass distribution \newcre{adopted in our simulations} is assumed to be that of a Singular Isothermal Ellipsoid (SIE, \citealt{KSB_SIE94}) perturbed by additional Gaussian Random Field (GRF) fluctuations \newcre{and an external shear}. An elliptical \cite{Sersic68} brightness profile is used to represent the lensed sources, and to which we add several small internal stellar structures (e.g., star-formation regions, satellite galaxies), described by circular S\'ersic profiles. For each background source, we extract magnitudes from the ``COSMOS" models provided by  the code~\textsc{Le Phare} (\citealt{Arnouts+99}; \citealt{Ilbert+06}) in order to simulate realistic \textit{gri}-composite images. \newcre{The lens and source parameters vary accordingly to the values in \Tab 1 of \cite{Petrillo2019}.}
\item[(2)] The non-lenses are a collection of $\sim12\,000$ galaxies from KiDS. This sample is comprised of a supersample of: (a) the same LRGs used for the \textit{lenses}; (b) randomly selected  galaxies from the survey with a \textit{r}-band
magnitude brighter than 21; (c) `false positives'  (e.g., mergers, ring galaxies, etc.) from earlier ConvNets; and (d) a sample of galaxies that were visually classified as spirals from an on-going GalaxyZoo project (\citealt{Willett2013}, Kelvin et al., in prep.).
\end{itemize}

\newcre{A more detailed description of the training sample preparation,} the results of the training phase, and a detailed discussion of the performance of the ConvNets are presented in \cite{Petrillo2019}.

\subsection{Application to the LRG sample}\label{SECresults}

\begin{figure}
\begin{center}
 {\includegraphics[width=85mm]{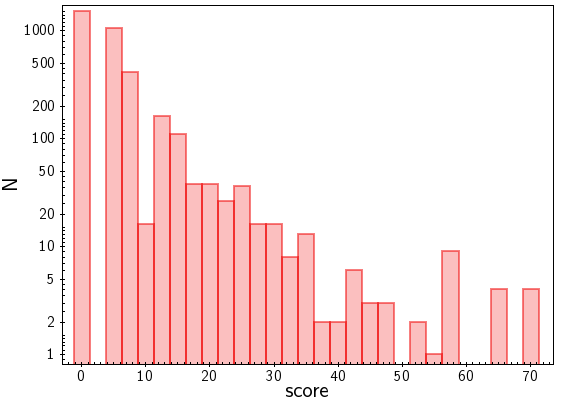}}
\caption{Histogram of the numerical rankings from the visual inspection of 3500 targets, selected by the ConvNets, by seven human classifiers. \newcre{See \Sec\ref{SECresults} for the detailed discussion of the results.}}
\label{FIGscores}
\end{center}
\end{figure}

\captionsetup[subfigure]{labelformat=empty}
\begin{figure*}
\begin{center}
\subfloat[70, 0.957, <0.8]{\includegraphics[width=40mm]{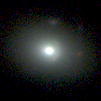}}
\subfloat[70, 1. , 1.]{\includegraphics[width=40mm]{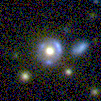}}
\subfloat[70, 0.999, 0.999]{\includegraphics[width=40mm]{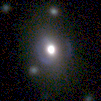}}
\subfloat[70, 1. , 0.999]{\includegraphics[width=40mm]{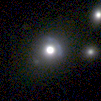}}
\\
\subfloat[64, 1., 1.]{\includegraphics[width=40mm]{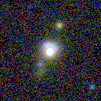}}
\subfloat[64, <0.8, 0.937]{\includegraphics[width=40mm]{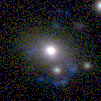}}
\subfloat[64, 0.887, 0.901]{\includegraphics[width=40mm]{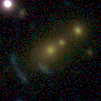}}
\subfloat[64, 0.989, 0.9]{\includegraphics[width=40mm]{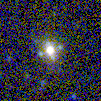}}
%\hfill
\caption{The candidates classified through visual inspection with the two topmost scores, 70 and 64. Below each image are shown the visual inspection score followed by the $p$-values of the 1-band and 3-band ConvNets. Each image has dimensions $20 \times 20$ arcseconds.}
\label{FIGmaxscoresvisisp}
\end{center}
\end{figure*}

The ConvNets described in the previous subsection are both applied to the LRG sample, and only targets with $p>0.8$ (returned from either of the ConvNets) are selected. This threshold is chosen to obtain a reasonable number of `true positives' and, at the same time, not contaminate the sample with a large number of `false positives'. \cite{Petrillo2019} present an extensive analysis of the performance of these ConvNets by choosing different $p$-value thresholds. With this threshold, the 3-band ConvNet picks 1689 candidates, while the one-band ConvNet picks 2510 candidates. These numbers correspond to fractions of $\sim 1.9$ and $\sim 2.8$ per cent of the LRG sample, respectively. We find a total of (exactly) 3500 unique candidates with $p>0.8$ since 699 galaxies are common between both ConvNets. We refer to this sample of 3500 unique targets as the \textit{ConvNet sample}.

By setting the threshold value $p$ to 0.8, however, we still expect the presence of many false positives in the \textit{ConvNet sample} ($\sim90$ per cent; \citealt{Petrillo2019}). 
To validate the candidates, selected by the ConvNets, we conduct a visual inspection: seven of the authors of this paper -- referred to as ``classifiers" --  examine the $101\times101$ pixels RGB composite image, created with STIFF\footnote{\href{http://www.astromatic.net/software/stiff}{\tt http://www.astromatic.net/software/stiff}} \citep{Bertin2012}. The classifiers have only three possible choices for each source being a lens: \textit{Sure}, \textit{Maybe}, and \textit{No lens}. We translate each of these categories into a numerical value in the same way as was done by \cite{Petrillo2017}:\\

\begin{tabular}{lll}
A: & \textit{Sure lens} & 10 points. \\
B: & \textit{Maybe lens}  & 4 points. \\
C: & \textit{No lens} &  0 points. \\
\end{tabular}
\vspace{4mm}

\begin{figure*}
\begin{center}
\subfloat[58, 0.999, <0.8]{\includegraphics[width=35mm]{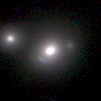}}
\subfloat[22, 0.939, <0.8]{\includegraphics[width=35mm]{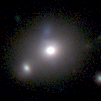}}
\subfloat[54, 1., 0.999]{\includegraphics[width=35mm]{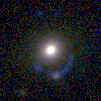}}
\subfloat[24, 0.999, 1.]{\includegraphics[width=35mm]{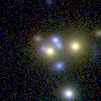}}
\subfloat[64, 0.989, 0.9]{\includegraphics[width=35mm]{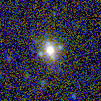}}
\\
\subfloat[]{\includegraphics[width=29mm]{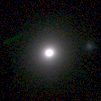}}
\subfloat[]{\includegraphics[width=29mm]{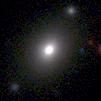}}
\subfloat[]{\includegraphics[width=29mm]{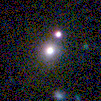}}
\subfloat[]{\includegraphics[width=29mm]{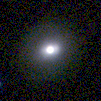}}
\subfloat[]{\includegraphics[width=29mm]{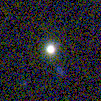}}
\subfloat[]{\includegraphics[width=29mm]{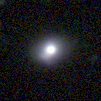}}
\caption{First row: images of 5 known confirmed lenses re-discovered by the ConvNets. Below each image are shown the visual inspection score followed by the $p$-values of the 1-band and 3-band ConvNets.  Second row: known lenses in the LRG sample not identified by the ConvNets. All the images have dimensions $20 \times 20$ arcseconds.}
\label{FIGknownlenses}
\end{center}
\end{figure*}

\noindent As a result, the maximum score that any one galaxy candidate can obtain is 70, i.e. when all human classifiers think it is surely a lens. A histogram with the numerical results of the visual inspection is shown in \Fig\ref{FIGscores}. About $\sim57$ per cent of the initial 3500 candidates selected by the ConvNets (i.e., 1983 candidates) have at least one classifier selecting it as a  \textit{Sure lens} or \textit{Maybe lens}. Only four candidates achieve the maximum score. \Fig\ref{FIGmaxscoresvisisp} presents the eight candidates that received the two highest scores, i.e., 64 and 70. Among them, there is one confirmed quad lens, J115252+004733 \newcre{(bottom right panel, \citealt{More2017})}. It is worth noting that, within the full \textit{ConvNet sample}, there are five confirmed lenses:  J114330-014427, J1025-0035 \citep{SLACS2008}, J085446-012137 \citep{Cabanac2007}, CSWA 5 \citep{Christensen2010} and J115252+004733 \citep{More2017} classified with scores of 58, 22, 54, 24, 64 and 64 respectively (see \Fig\ref{FIGknownlenses}). 
Naturally this means that none of these confirmed lenses were flagged as \textit{Sure lens} by all classifiers. However, these sources are often confirmed as lensed through high angular resolution HST (Hubble Space Telescope) follow-up, which makes it unsurprising that \newcre{they are not classified as secure lenses} in ground-based KiDS data.
In the LRG sample there are other six known gravitational lenses which have not been identified by our ConvNets. However, the KiDS images of these objects do not exhibit striking lensing features and, thus, they are hardly recognizable as strong lenses.

\captionsetup[subfigure]{labelformat=empty}

\begin{figure}
\begin{center}

\subfloat[4]{\includegraphics[width=40mm]{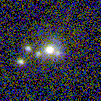}}%\hspace{\fill}
\subfloat[42]{\includegraphics[width=40mm]{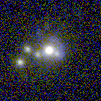}}
\caption{Images of the same candidate retrieved by the ConvNets in two different survey tiles. The scores from the visual classification (the numbers below the images) are different because of the different quality of the images. Each image has dimensions $20 \times 20$ arcseconds.}
\label{FIGdoppiaim}
\end{center}
\end{figure}

\captionsetup[subfigure]{labelformat=empty}
\begin{figure*}
\begin{center}
\subfloat[0]{\includegraphics[width=35mm]{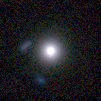}}
\subfloat[12]{\includegraphics[width=35mm]{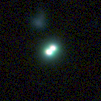}}
\subfloat[4]{\includegraphics[width=35mm]{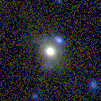}}
\subfloat[58]{\includegraphics[width=35mm]{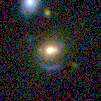}}
\subfloat[4]{\includegraphics[width=35mm]{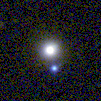}}
\\
\subfloat[36]{\includegraphics[width=35mm]{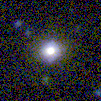}}
\subfloat[42]{\includegraphics[width=35mm]{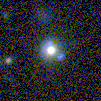}}
\subfloat[18]{\includegraphics[width=35mm]{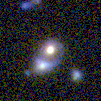}}
\subfloat[12]{\includegraphics[width=35mm]{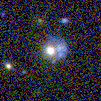}}
\subfloat[30]{\includegraphics[width=35mm]{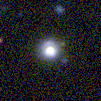}}
\\
\subfloat[28]{\includegraphics[width=35mm]{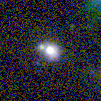}} %#HSC
\subfloat[52]{\includegraphics[width=35mm]{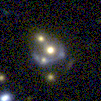}}
\subfloat[70]{\includegraphics[width=35mm]{2854.png}}
\subfloat[10]{\includegraphics[width=35mm]{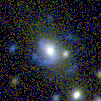}}
\subfloat[20]{\includegraphics[width=35mm]{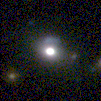}}
\\
\subfloat[18]{\includegraphics[width=35mm]{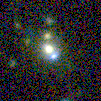}}
\subfloat[30]{\includegraphics[width=35mm]{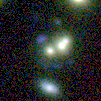}}
\subfloat[64]{\includegraphics[width=35mm]{2729.png}}
\subfloat[24]{\includegraphics[width=35mm]{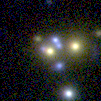}}
\subfloat[24]{\includegraphics[width=35mm]{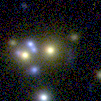}}
%\hfill
\caption{Candidates selected by the 3-band ConvNet with $p>0.999$. The scores from the visual inspection are shown below the images. \newcre{In the last row, the lens J1244+0106 is shown} twice because it appears twice in the LRG sample since is centred on two different LRGs. Each image has dimensions of $20 \times 20$ arcseconds.}
\label{FIGmaxscores3bands}
\end{center}
\end{figure*}

\captionsetup[subfigure]{labelformat=empty}
\begin{figure}
\begin{center}
\subfloat[32]{\includegraphics[trim={5mm 5mm 5mm 5mm},clip,width=25mm]{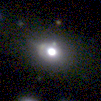}}
\subfloat[28]{\includegraphics[trim={5mm 5mm 5mm 5mm},clip,width=25mm]{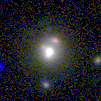}}
\subfloat[30]{\includegraphics[trim={5mm 5mm 5mm 5mm},clip,width=25mm]{2967.png}}
\\
\subfloat[52]{\includegraphics[trim={5mm 5mm 5mm 5mm},clip,width=25mm]{3154.png}}
\subfloat[70]{\includegraphics[trim={5mm 5mm 5mm 5mm},clip,width=25mm]{2854.png}}
\subfloat[64]{\includegraphics[trim={5mm 5mm 5mm 5mm},clip,width=25mm]{2729.png}}
\\
\subfloat[54]{\includegraphics[trim={5mm 5mm 5mm 5mm},clip,width=25mm]{2558.png}}
\subfloat[70]{\includegraphics[trim={5mm 5mm 5mm 5mm},clip,width=25mm]{0805.png}}
\subfloat[28]{\includegraphics[trim={5mm 5mm 5mm 5mm},clip,width=25mm]{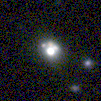}}
%\hfill
\caption{Candidates selected by the 1-band ConvNet with $p>0.999$. The scores from the visual inspection are shown below each image. Each image has dimensions $14 \times 14$ arcseconds.}
\label{FIGmaxscores1band}
\end{center}
\end{figure}

The visual classification appears to depend on the signal to noise ratio. For example, the candidate SCJ083726+015639, found in HSC data by \cite{Sonnenfeld2018}, is present in two adjacent KiDS tiles, and the ConvNets retrieve it from both tiles (the ConvNets select three more HSC candidates). Nevertheless, the human classifiers, in general, give very different scores to the same candidate depending on the quality of the images (\Fig\ref{FIGdoppiaim}). Thus, it is fair to assume that many `good' candidates are lost from our sample if we preferentially select only those candidates with high visual-inspection score. On the other hand, there are also clearly cases where the ConvNets select candidates without any human-identifiable lensing feature being present. 

\begin{figure}
\begin{center}
 {\includegraphics[width=90mm]{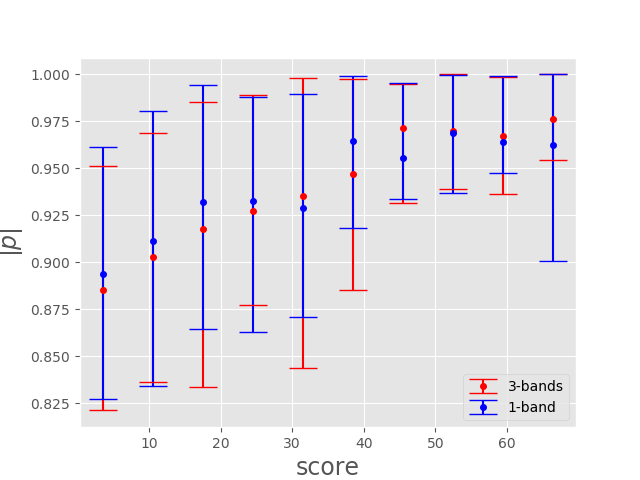}}
\caption{Average of the $p$-values given by the two ConvNets divided in bins of the score of the visual inspection. The vertical bars correspond to the 16-84 percentile of the distributions.}
\label{FIGp-score}
\end{center}
\end{figure}

To examine the other extreme of the classification, \Figs\ref{FIGmaxscores3bands} and \ref{FIGmaxscores1band} present the candidates that the ConvNets classify with values of $p>0.999$, along with the scores from our visual inspection. For the 3-band ConvNet, some of these extremely high-confidence ConvNet candidates received low visual classification scores; there is even a case with visual-inspection score of zero.
It is clear that there remains significant disagreements between human and ConvNet classifications, and that both classification methods are prone to some level of bias and error. Nonetheless, \Fig\ref{FIGp-score} demonstrates that the visual-inspection scores and the $p$-values are indeed correlated. The figure shows a positive correlation between average values of the $p$-values for different bins of the visual inspection score.
Hence, even if the classification schemes from humans and ConvNets differ, both tend to agree to a certain extent on what constitutes a `good' lens candidate.  

\begin{figure*}
\begin{center}
 {\includegraphics[width=87mm]{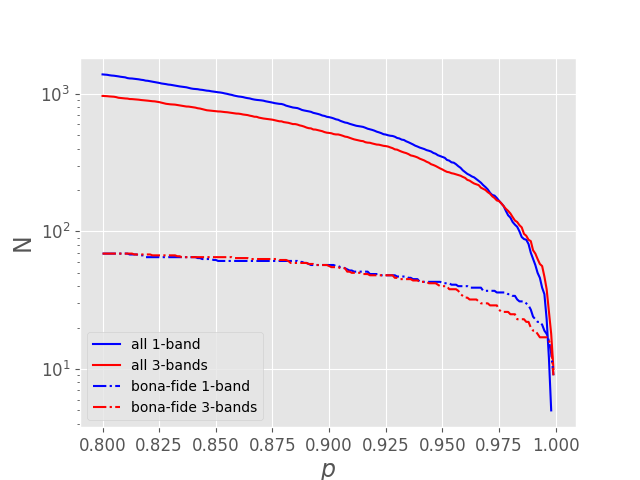}}
 {\includegraphics[width=87mm]{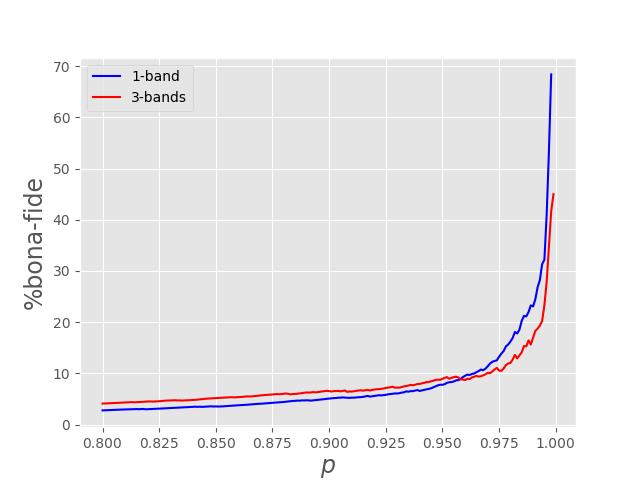}}
\caption{On the left we plot the number of targets retrieved by the two ConvNets and the number of ``bona fide" candidates as a function of the threshold of detection $p$. 
On the right we show the percentage with respect to the total number of retrieved candidates as a function of the threshold of detection $p$.}
\label{FIGpanalysis}
\end{center}
\end{figure*}

Even if there is no obvious inspection-score below which the candidates are no longer reliable, we nonetheless observe an increase in the fraction of good candidates with increasing score. Therefore by defining some fiducial threshold for the visual inspection score, above which one considers the targets as reliable candidates, we can investigate how the number of retrieved candidates (and the degree of contamination) vary as a function of the threshold set on the value of $p$. \Fig\ref{FIGpanalysis} presents these correlations for all the ConvNet candidates and for a ``bona fide'' sub-sample composed of targets with a visual inspection score $\geqslant28$. This is a fiducial value of the score which corresponds to a) \textit{maybe lens} given by all the classifiers or to b) \textit{sure lens} given by two classifiers and \textit{maybe lens} from other two classifiers. \newcre{In particular, in the left panel of \Fig\ref{FIGpanalysis} we see how the number of retrieved candidates changes as a function of the value of $p$, greatly decreasing when $p$ is approaching to 1. This change is more gentle in the case of the ``bona fide" sample. The right panel shows that the fraction of ``bona fide" systems is increasing with $p$, reaching the lowest contamination degree when $p$ is close to 1. This latter result confirms the correlation among the visual inspection score and $p$, previously shown in \Fig\ref{FIGp-score}.} 

\section{The LinKS sample candidates}\label{SEClinks}
We define the ``LinKS (Lenses in the Kilo-Degree Survey) sample'' as the full sample of 1983 gravitational lens candidates retrieved with $p>0.8$ and a score from the visual inspection greater than zero. The sample contains five previously confirmed strong lenses (see \Fig\ref{FIGknownlenses}; \citealt{Cabanac2007, SLACS2008, Christensen2010, More2017}) and 12 lens candidates discovered in the HSC data \citep{Sonnenfeld2018,Wong2018}. 
This sample also contains the ``bona fide" subsample, composed of the 89 candidates which have a visual inspection score $\geqslant28$, which we defined in Sect.\ref{SECresults}. We note that by relaxing this inspection score requirement further, for example to $\geqslant16$ (i.e. the score corresponding to four \textit{maybe a lens}), we are able to produce a subsample of 308 candidates. Nonetheless, we opt to define our ``bona fide" sample using the more stringent $\geqslant28$ requirement.

Information about the data products provided for the LinKS sample, along with images for each of the 89 ``bone fide'' candidates, is provided in Appendix \ref{SECtoplinks}. Additional information is also provided at the \newcre{the LinKS webpage}\footnote{\href{http://www.astro.rug.nl/lensesinkids}{\tt http://www.astro.rug.nl/lensesinkids }}. 

\subsection{Candidate properties}\label{SECpopprop}
\begin{figure*}
\begin{center}
\includegraphics[width=80mm]{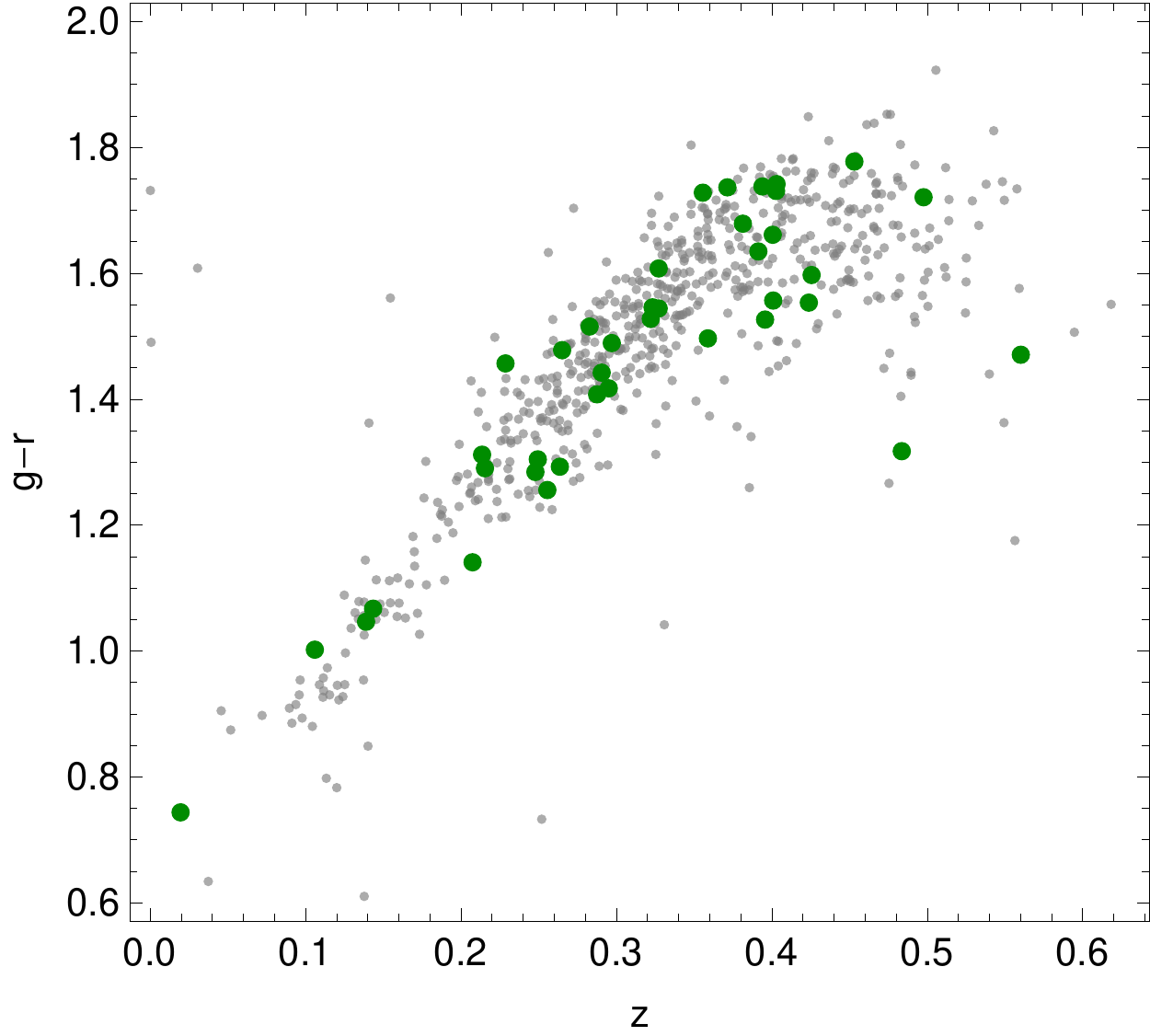}
\includegraphics[width=82mm]{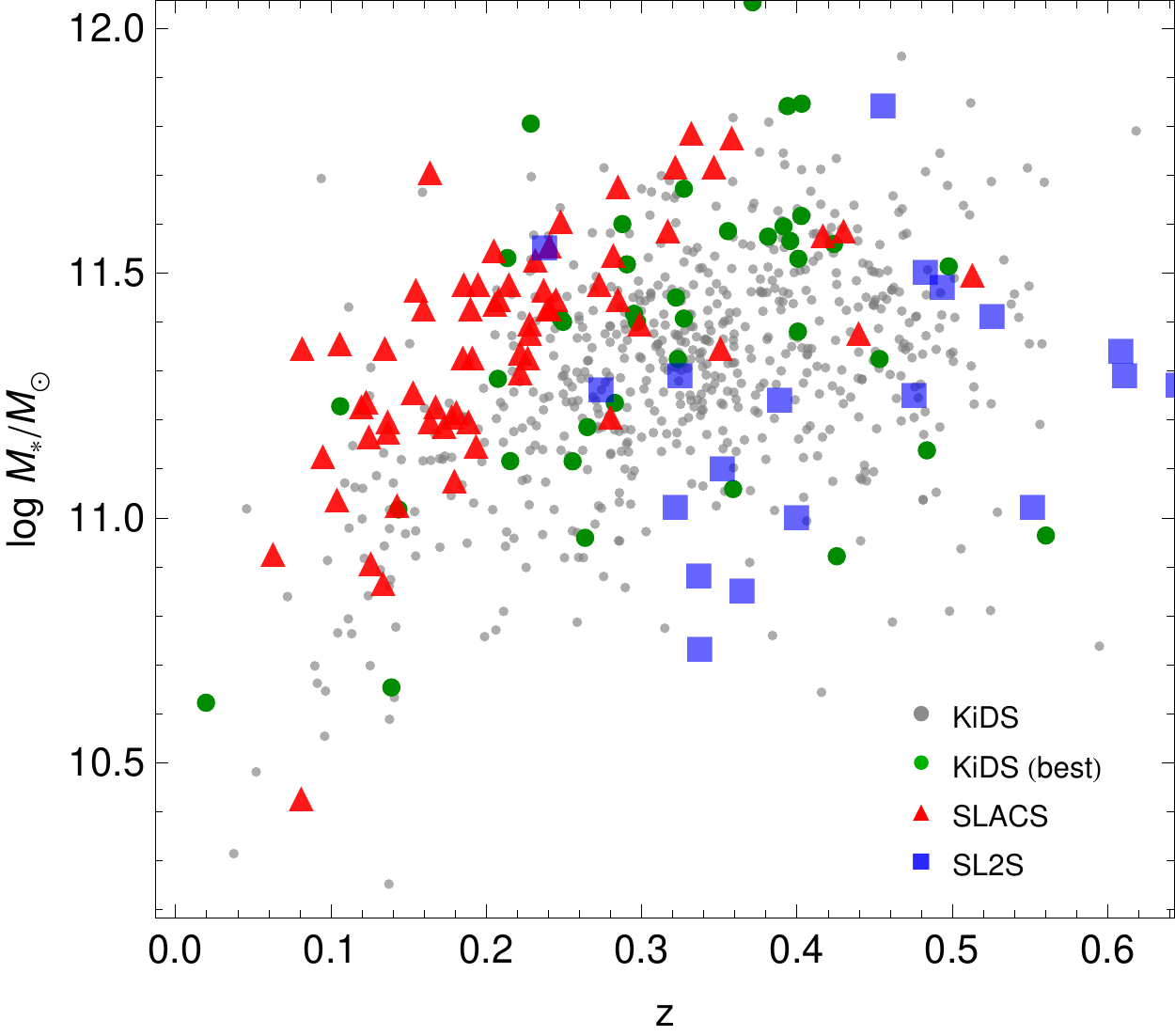}
\caption{The $g-r$ observer-frame colour, corrected for Galactic extinction (left panel) and stellar mass (right panel) versus redshift for a subsample of 659 ConvNet candidates with spectroscopic redshift available (grey dots). The subsample of best candidates with a visual score larger than 28 are shown as green points. Stellar masses (see \Sec\ref{SECpopprop}) are compared with the SLACS sample from \citet[red triangles]{Auger+09_SLACSIX} and the SL2S sample from \citet[blue squares]{Sonnenfeld+13_SL2SIV}.}
\label{FIGmasses}
\end{center}
\end{figure*}

In this section we summarise the main characteristics of the LinKS sample. To enable this analysis, we rely on candidates with known spectroscopic redshift publicly available from SDSS DR14, GAMA DR3 and 2dFLenS \citep{Abolfathi2018,Baldry2018,Blake+16_2dflens}. 
We also incorporate accurate multi-band colours as measured by the Gaussian Aperture and PSF (GAaP) code.
Briefly, GAaP produces fluxes measured in Gaussian-weighted apertures, which are modified per-source and per-image, so as to produce seeing-independent estimates flux estimates across different observations/bands. The aperture modification calculation requires that the PSF of the image be both homogeneous and Gaussian, and so prior to running GAaP each survey tile has its PSF Gaussianised over the full field of view.  Importantly, GAaP magnitudes are not total, and preferentially weight the central, redder parts of our lens galaxies. This acts to reduce the contamination of the outer (blue) features of the lens candidates (i.e. the lensed arcs), and improve the fidelity of lens-candidate SED models. 
In this section we have chosen to limit our analysis to the LinKS sample in the KiDS-North patch\footnote{The fourth KiDS data release consists of multi-band GAaP catalogues for both the Northern and Southern patches, but we chose to limit our analysis to a preliminary set of 497 tiles that were processed by Astro-WISE at the start of this analysis. We note that some improvements have been made to the GAaP catalogues during the course of this work, in particular the calibration of the {\it u}-band zero-points has been refined. We do not expect these updates to significantly impact our conclusions.}. This selection reduces our LinKS sample to 659 candidates, of which 41 (out of 89) are in the ``bona fide" subsample.
We show the observer-frame $g-r$ colour in terms of redshift of these candidates in the left panel of \Fig\ref{FIGmasses}. Due to our initial selection criteria (see \Sec\ref{sec:LRG}) all of our candidates exhibit red colours, with $g-r \sim 0.8$ at $z\sim0$ and $g-r \sim 1.7$ at the highest redshifts $z \sim 0.5$. Visually the ``bone fide'' candidates seem to sample the colour distribution of the entire sample without bias; they are otherwise unexemplary. To further characterize the sample of candidates, and allow for a comparison with the literature, we then estimate stellar masses for the subsample of our sources with spectroscopic redshifts\footnote{Robust stellar masses are available from the literature for those KiDS galaxies that are also contained in SDSS and GAMA. However, in order to have homogeneous results for all the candidates, we determine the masses for the whole sample using KiDS 4-band photometry.}. Following \cite{Petrillo2017}, we estimate stellar masses using the software \textsc{Le Phare}
(\citealt{Arnouts+99}; \citealt{Ilbert+06}), which does a $\chi^{2}$ fitting between colours from stellar population synthesis (SPS) models and the observed colours. We employ single burst SPS models from \citet[BC03]{BC03} and a \cite{Chabrier01} IMF, allowing the stellar population age to vary and assuming metallicities in the range (0.005--2.5 \Zsun). The maximum age is set by the age of the Universe at the redshift of the galaxy, with a maximum value at $z=0$ of $13\, \rm Gyr$. We do not consider internal extinction, and our models assume zero redshift uncertainty. We adopt the GAaP $ugri$ magnitudes \maggaap\ and related $1\, \sigma$ uncertainties (Kuijken et al. 2018, in prep.), corrected for Galactic extinction using the map by \cite{Schlafly_Finkbeiner11}. The \textit{r}-band \Mauto\ is used to correct the results of \textsc{Le Phare} for missing flux\footnote{GAaP magnitudes do not trace the whole galaxy light distribution, for this reason we need to correct the stellar masses $\log M_{\rm \star}^{\textsc{Le Phare}}$ for missing flux, using the following formula $\log M_{\rm \star} = \log M_{\rm \star}^{\textsc{Le Phare}} + 0.4*(\maggaapr-\Mautor)$. This could contaminate the lens mass estimate but, since the lensed sources are usually blue, the impact on \Mautor\ is usually small.}.
The typical uncertainty on the stellar mass estimates (provided by {\textsc LePhare}) is $\sim 0.1-0.2$ dex. Stellar masses are shown as a function of redshift in \Fig\ref{FIGmasses}, and compared with SLACS (\citealt{Auger+09_SLACSIX}) and SL2S (\citealt{Sonnenfeld+13_SL2SIV}) data. \newcre{Consistently with \cite{Petrillo2017},} the selected candidates have redshifts in the window $0.1 \lsim z \lsim 0.5$, with a median value of $0.33$, while the stellar masses are typically larger than $10^{11}\Msun$, with an average value of $\sim 2 \times 10^{11}\Msun$. We note, of course, that the choice of IMF significantly influences the final mass estimates; using a \cite{Salpeter1955} IMF instead of a Chabrier IMF causes inferred stellar masses to increase by a factor of $\sim2$ with no change to observed colours \citep{Tortora+09}. The ``bone fide'' candidates are shown in green in both panels. They span a similar range of redshifts and masses as the whole sample, with a marginal indication that they may preferentially sample higher stellar masses. 

\subsection{Predictions and Prospects: Euclid and LSST}\label{SEClenspop}

Using the \textsc{LensPop} code presented in \cite{collet2015},  \cite{Petrillo2017} estimated  that the maximally retrievable number of strong lens candidates in a fully complete KiDS survey would be $\sim2400$. For a $\sim900$ square degrees area such as that considered in this paper, ignoring the masked area of the survey, we would there expect to find $\sim1700$ possible strong lenses. If we further consider only those lenses that satisfy our LRG colour-magnitude cuts (\Sec\ref{sec:LRG}), and which have an Einstein radius larger than one arcsecond (i.e. the range on which the ConvNets have been trained; \newcre{see \Tab 1 in \citealt{Petrillo2019}),} this number reduces further to about $\sim450$ retrievable strong lenses. Their average distribution in redshift is consistent with the actual distribution of our retrieved candidates of the previous subsection, peaking at a value of $z\sim0.3$. Our samples here therefore fully encompass the predicted $\sim450$ retrievable strong lenses from \textsc{LensPop}: the full sample of LinKS candidates containing $\sim4\times$ the number of predicted sources, and the bone fide sample containing $\sim5\times$ too few. We note again, though, that by relaxing the visual inspection score requirement to, e.g., $\geqslant16$ (the score corresponding to four \textit{maybe a lens}) one can create a wider ``bone fide'' sample containing 308 candidates;  $\sim68$ per cent of the \newcre{retrievable lenses predicted by \textsc{LensPop}}.
%
%, a number similar to the theoretical prediction from %\textsc{LensPop}. 
%
Nonetheless, we continue to conservatively consider only the 89 sources in our ``bona fide" subsample to be genuine lenses, and conclude that this sample is complete at the level of  $\sim20$ per cent. 

\newcre{In the following, we predict the number of lenses expected in future surveys utilising the depth and breadth of the future Euclid and LSST surveys, and the performance of our ConvNets in retrieving strong lenses within these future datasets.}

\paragraph*{Euclid.} \cite{collet2015} predicts that there will be $\sim 170\,000$ potential lenses in Euclid. \cite{Petrillo2019} extended this analysis by estimating the number of lenses with an Einstein radius larger than 1 arcsecond and with a redshift $z < 0.5$, which roughly corresponds to our LRG colour cut selection. This reduces the number of potential strong lenses  to $\sim20\,000$ in the $15\,000$ square degrees of the completed survey.  With the same strategy used in this paper, we conservatively estimate that between $\sim5000$ and $\sim15\,000$ lenses will be retrievable \newcre{with ConvNets} from the completed Euclid survey. These numbers assume that the 1-band ConvNet performs at least as well on Euclid data as it does on KiDS data, in the same parameter-domain, and that it is possible to pre-select LRGs with the aid of ground-based multi-band observations and the IR-bands from Euclid. We note, though, that Euclid data will have better image quality than KiDS, which will allow the training of more effective algorithms over a wider parameter space. Furthermore, it will allow improved recognition and rejection of false positives via visual inspection. These considerations all lead to our assessment that our estimate of the number of retrievable strong lenses is conservative.  

\paragraph*{LSST.} The above forecast can also be performed for LSST, and moreover with greater accuracy, as LSST will observe in the same $g$, $r$, and $i$ filters as does KiDS. We find that the number of potentially discoverable lenses in LSST, with an Einstein radius larger than one arcsecond and with our invoked LRG colour selection, is $\sim20\,000$ over the 20\,000 square degrees of the completed survey. Therefore, as in Euclid, we estimate that between $\sim5000$ and $\sim15\,000$ lenses may be retrievable from the completed LSST survey data with our ConvNets.
%Moreover, from \Fig9 of \cite{Petrillo2019} we expect that the \textit{ConvNet sample} (3500 targets) it is roughly $90\%$ contaminated (consistently with the result of \Fig\ref{FIGpanalysis} of this paper). This translates to $\sim350$ expected good candidates in the sample, that is consistent with the theoretical prediction obtained with \textsc{LensPop}. 

%\begin{figure}
%\begin{center}
% {\includegraphics[width=85mm]{LRG_distribution.png}}
%\caption{\textit{g-r} colour-redshift distribution of the LRG sample (green dots; \Sec\ref{sec:LRG}) and the LinKS sample (blue dots; \Sec\ref{SECresults}). The BPZ photometric redshift is plotted, except for the candidates with an available spectroscopic redshift.}
%\label{FIGLRGdistribution}
%\end{center}
%\end{figure}

%\begin{figure}
%\begin{center}
%{\includegraphics[width=85mm]{z_phot_vs_z_spec.png}}
%\caption{Spectroscopic vs photometric redshift for 430 candidates for which both the redshifts are available.}
%\label{FIGzspec_zphot}
%\end{center}
%\end{figure}

\section{The full sample candidates}\label{SECfullsurvey}

Visual inspection of strong lens candidates selected by the ConvNets is a time-consuming task. However investing such time to achieve increased purity and completeness of the recovered candidate sample is worth the effort. But lowering the $p$-value threshold above which lens candidates are defined, or significantly increasing the survey area (and thus significantly increasing the absolute number of $p\geqslant0.8$ candidates) naturally only exacerbates this task. As such, performing the visual inspections completed here for much larger target samples, such as those expected \newcre{from Euclid and LSST}, will likely be prohibitive. In these cases, one may want to reduce the number of candidates to visually inspect by increasing the $p$-threshold required for candidacy definition. However it is unclear how such an increase may influence the number of lens-candidates returned. Furthermore, if the scientific aim is to establish a complete strong-lens sample that is unbiased in its lens properties, then such a high threshold may be counter-productive. The \textit{LRG sample} used in this paper is a distinct sub-sample of massive Early Type Galaxies (ETGs) which lack (active) star formation and therefore have profiles which allow easier separation of foreground lenses from lensed images, which are often blue star-forming galaxies, as demonstrated in SLACS. In this work we use the LRG sample because we expect most of the lenses to be massive ETGs. However, selecting such a sample of galaxies is not always straightforward and can lead to the loss of potential lenses; LRGs do not represent the entire population of galaxies and hence the entire strong-lensing cross-section. 

For this reason, it is interesting to see how the ConvNets perform on a less restricted and much larger sample of galaxies. We explore these issues in \Sec\ref{subsec:high-purity-KiDS} by applying the ConvNets to the \textit{full sample}, but with a higher threshold in $p$, in order to reduce the visual inspection effort. In \Sec\ref{subsec:high-purity-Euclis-LSST}, we then translate the outcome to the planned Euclid and LSST surveys and analyse the advantages and applicability of such a strategy. Finally, in \Sec\ref{SECbonus} we present a composite sample of lens candidates collected from various ConvNets, applied to the \textit{full sample}, that were run during the ConvNet optimisation process. Each of these runs was less efficient than the final ConvNets employed in the main body of this work, but sometimes yielded distinct lenses which we have subsequently collated.

\subsection{A high-purity sample}\label{subsec:high-purity-KiDS}

\begin{figure*}
%\begin{center}
\subfloat{\includegraphics[width=30mm]{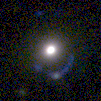}}
\subfloat{\includegraphics[width=30mm]{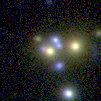}}
\subfloat{\includegraphics[width=30mm]{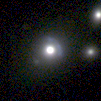}}
\subfloat{\includegraphics[width=30mm]{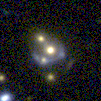}}
\subfloat{\includegraphics[width=30mm]{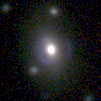}}
\subfloat{\includegraphics[width=30mm]{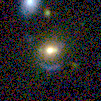}}
\\
\vspace{-11pt}
\subfloat{\includegraphics[width=30mm]{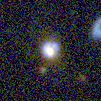}}
\subfloat{\includegraphics[width=30mm]{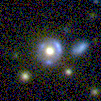}}
\subfloat{\includegraphics[width=30mm]{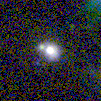}}
\subfloat{\includegraphics[width=30mm]{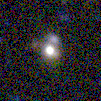}}
\subfloat{\includegraphics[width=30mm]{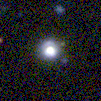}}
\subfloat{\includegraphics[width=30mm]{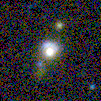}}
\\
\subfloat{\includegraphics[width=30mm]{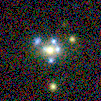}}
\subfloat{\includegraphics[width=30mm]{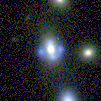}}
\subfloat{\includegraphics[width=30mm]{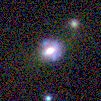}}
\subfloat{\includegraphics[width=30mm]{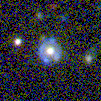}}
\subfloat{\includegraphics[width=30mm]{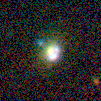}}
\subfloat{\includegraphics[width=30mm]{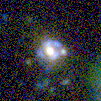}}
\vspace{-11pt}
\\  
\subfloat{\includegraphics[width=30mm]{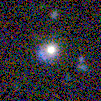}}
\subfloat{\includegraphics[width=30mm]{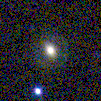}}
\subfloat{\includegraphics[width=30mm]{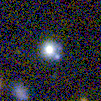}}
\subfloat{\includegraphics[width=30mm]{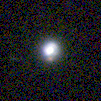}}       
\subfloat{\includegraphics[width=30mm]{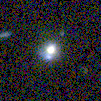}}      
\subfloat{\includegraphics[width=30mm]{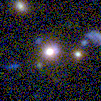}}
\vspace{-11pt}
\\
\subfloat{\includegraphics[width=30mm]{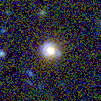}}
\subfloat{\includegraphics[width=30mm]{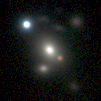}}       
\subfloat{\includegraphics[width=30mm]{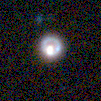}}
\subfloat{\includegraphics[width=30mm]{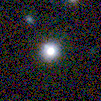}}
\subfloat{\includegraphics[width=30mm]{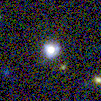}}
\subfloat{\includegraphics[width=30mm]{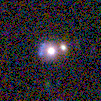}}
\caption{Images of the sample of candidates retrieved running the two ConvNets on the \textit{full sample}, averaging the predictions and selecting those with $p>0.999$. This sample is $>90$ per cent pure and requires very little human intervention. The upper block of 12 galaxies is part of the \textit{LRG sample} while the bottom 18 galaxies are exclusively part of the \textit{full sample}. More information on the latter candidates can be found in the Appendix. Each image has dimensions $20 \times 20$ arcseconds.}
\label{FIGmixedfullsurvey}
%\end{center}
\end{figure*}

We run the two ConvNets on the \textit{full sample} (930\,651 galaxies) rather than on the smaller but purer \textit{LRG sample} (88\,327 galaxies). To obtain a sample of lens candidates that is both pure and limited in size, and in order to reduce the visual inspection load, we average the predictions from both ConvNets into a single predictive parameter $p$. We select candidates with an average value of $p$ larger than $0.999$. With this selection we obtain just 30 strong lens candidates (\Fig\ref{FIGmixedfullsurvey}); $0.003$ per cent of the \textit{full sample}. 
%This sample is expected to have an extremely high purity (for lens-candidate searches) of $\sim93$ per cent. 
When visually inspected, we find that this sample is extremely pure and, more in particular, it is composed of\footnote{\newcre{This sample has been visually inspected using a classification scheme similar to, but not the same as, the one adopted for the LinKS sample. For sake of brevity we omit details about this classification.}}:  
\begin{itemize}
\item 2 confirmed lenses \citep{Cabanac2007, SLACS2008};
\item 1 candidate discovered by \cite{Sonnenfeld2018};
\item 1 quad recently identified by \cite{Sergeyev2018};
\item 14 very-likely genuine lenses;
\item 10 potential lenses;
\item 2 possible contaminants. 
\end{itemize}

\noindent This result attests to the capability of the ConvNets to find lens candidates in a sample slightly different from what it was trained on. We note that 18 of the 30 candidates retrieved in this manner are not part of the \textit{LinKS sample} because they did not satisfy the LRG cut in \Sec\ref{sec:LRG} (see \Sec\ref{SECbonus} for more information on these candidates). We note further that it is entirely possible that some of these candidates fail our LRG colour-magnitude selection explicitly because of contamination by the bright blue lensing features that we are attempting to locate; a clear drawback of such an LRG selection with imperfect photometry.

\captionsetup[subfigure]{labelformat=empty}

\subsection{Small high-purity Euclid \& LSST samples}\label{subsec:high-purity-Euclis-LSST}

Considering that, theoretically, the number of recoverable lenses in $\sim900$ square degrees of KiDS is at most $\sim1700$ (see \Sec\ref{SEClenspop}), our recovery of only 30 candidates in Sect.~\ref{subsec:high-purity-KiDS} implies that a $p>0.999$ setup will only recover $\sim2$ per cent of possibly retrievable lenses. If we turn this efficiency into a forecast for the 170\,000 total retrievable lenses in the full Euclid survey as predicted by \cite{collet2015}, we expect to find $\sim3000$ candidates with a $>90$ per cent purity which are retrievable with  minimal human intervention. Such a sample represents the often called ``low-hanging fruit" of strong lenses within Euclid, as these sources are expected to occupy a limited but easily accessible part of parameter space (i.e.,\ large Einstein radii and low redshifts). Note again that we expect this number to be conservative, as with our other forecasts presented in \Sec\ref{SEClenspop}, as Euclid lenses will be observed with a much higher angular resolution than KiDS lenses, and will be detected with ConvNets trained on higher fidelity data. 
Near-infrared colours will also help to down-select lens candidates since, being less sensitive to the dust and mapping a wider wavelength baseline, they will provide a more efficient way to separate LRGs from star forming galaxies.

Nonetheless, even in this conservative case, the number of lenses forecast here would be one to two orders of magnitude larger than the number detected in any previous or ongoing strong lens survey. Finally, as in \Sec\ref{SEClenspop}, a similar number of easy candidates may be expected from LSST surveys.

\subsection{The ``bonus sample"}\label{SECbonus}

The sample presented in this section includes 200 strong lens candidates discovered serendipitously during previous ConvNet runs that are not part of the LinKS sample. The candidates in this \textit{Bonus sample} have not gone through the same rigorous visual inspection as those in the LinKS sample, and subsequently cannot be considered to be as statistically well defined. However if we apply the ConvNets to these candidates with a threshold $p>0.8$, 160 candidates pass this threshold in at least one of the two ConvNets, i.e., $80$ per cent of the sample. Detailed data related to this sample can be found online\footnote{\href{http://www.astro.rug.nl/lensesinkids}{\tt http://www.astro.rug.nl/lensesinkids }} (see the Appendix). This sample contains eight HSC survey lens candidates \citep{Sonnenfeld2018} and four confirmed lenses J1452-0058 \citep{SLACS2008}, J142449-005322 \citep{Tanaka2016}, J010127-334319 \citep{Bettinelli2016} and KiDS0239-3211 \citep{Sergeyev2018}.

%\begin{figure}
%\begin{center}
%{\includegraphics[width=87mm]{dist_p_left.png}}
%\caption{Distribution of the $p$-values obtained by running the two ConvNets on the \textit{Bonus sample} (see \Sec\ref{SECbonus}). }
%\label{FIGdist_p_left}
%\end{center}
%\end{figure}

\section{Discussion and Conclusions}\label{SECdiscussion}

In this paper, we present several samples of lens candidates from the Kilo-Degree Survey (KiDS) which likely contain several hundred strong gravitational lenses.  To generate these samples, we apply two new lens-finder algorithms -- based on Convolutional Neural Networks (ConvNets) -- to a sample of $88,327$ LRGs, selected via a colour-magnitude cut, from 904 one-square-degree tiles of KiDS data. We visually inspect the candidates selected by these ConvNets and conservatively select 1983 rank-ordered candidates, which we designate the \textit{LinKS sample} (see \Sec\ref{SEClinks}). We further subset the data into subsamples of 219 more plausible candidates, and 89 highly likely candidates.  

We did not attempt to achieve a high level of statistical completeness in the samples of LRGs, nor in the samples of resulting lens candidates. We aimed instead to both maximise the number of lens candidates while minimising the fraction of false positives. Our colour-magnitude selection (\Sec\ref{sec:LRG}) aimed at choosing a large sample of massive (early-type) galaxies while specifically avoiding star-forming (e.g. spiral) galaxies and other contaminants. We note that \cite{Vakili2018} recently selected LRGs from KiDS data using the LRG colour-magnitude relation, and also computed their photometric redshifts; we anticipate that this sample could be utilised to compile a more statistically complete sample of KiDS LRG lens candidates in the future. In addition to the \textit{LinKS sample}, in \Sec\ref{SECbonus} we presented a \textit{Bonus Sample} that consists of two-hundred lens candidates. These lenses were serendipitously discovered in KiDS data, e.g., during previous experiments with various ConvNets. While these sources have not been rigorously scrutinised in the same manner as our main LinKS sample, and we therefore do not consider it to be as statistically well-defined as our main sample, it nonetheless contains a number of interesting strong lens candidates for future follow-up.  

From our KiDS strong lens candidates (together with those found by \citealt{Hartley2017} and \citealt{Spiniello2018}), the $\sim600$ \newcre{galaxy-scale} lens candidates found in DES \citep{Diehl2017,Jacobs2018,Spiniello2019} and HSC data  \citep{Sonnenfeld2018,Wong2018}, it will soon be possible to select a sample of confirmed lenses similar in size to the total number of gravitational lenses known today. For example the Masterlens database\footnote{\href{http://masterlens.astro.utah.edu/}{\tt http://masterlens.astro.utah.edu/}}, which assembles information on all known gravitational lenses, contains a total of $\sim600$ gravitational lenses discovered up to 2016. It is possible that the total number could be, by now, up to $\sim1000$ confirmed lenses and lens candidates. We believe it likely that strong lens searches within the KiDS, DES, and HSC surveys could easily double this number -- accumulated over many decades -- within the next few years.

Despite the already considerable numbers of new lens candidates from KiDS, there are still many lens candidates to be discovered, especially in that part of parameter space that we have not, or rather not thoroughly, explored. In addition, the completed KiDS survey will cover an area of 1350 square degrees. We plan to apply our method to these completed KiDS data, together with that of \cite{Spiniello2018}, to find lensed quasars. Applying other complementary methods as \cite{Hartley2017} SVM will aid in maximizing the exploration of the parameter space.

Besides the LRG-selected sample, we have shown that is possible to tune the ConvNets to yield a  sample of lens candidates with considerable purity by using many more targets (i.e.\ about ten times more). In particular, we ran the lens-finders on a sample composed of 930\,651 galaxies (not just LRGs) and retrieved a sample of 30 strong lens candidates with an expected purity of $>90$ per cent. By selecting lens candidates in this way, we are able to considerably diminish the visual inspection load, although at the price of losing many genuine lenses. With a similar setup, though, it would be feasible to retrieve $\sim3000$ lens candidates from the future Euclid data set with minimal human intervention. \newcre{A similar number would be found by LSST.}

All these results can be enhanced further, especially by training the ConvNets with more complete training sets \citep{Petrillo2019}. In addition, a collection of genuine lens candidates, even in modest numbers, should allow one to fine-tune ConvNet lens-finders further to improve their classification capacity \citep{Tuccillo2018, Dom2018}. New gravitational lenses can also be used as training sets for future crowdsourced searches \citep{Marshall2016}. Finally, the candidates identified in this paper could be used to build a benchmark against which different lens-finders can be tested and compared, similar to analyses done with simulated data (e.g., \citealt{Metcalf2018})

Our results are very encouraging in light of future strong-lens surveys (e.g. \newcre{those utilising} Euclid and LSST) for which a naive strategy of visually inspecting galaxies to select lens candidates is entirely infeasible, given the enormous number of galaxies these new instruments will uncover. One can expect to compile samples of strong lenses from Euclid and LSST that are between one to two orders of magnitude larger than the samples compiled by any survey to date, and with minimal human effort. 

\section*{Acknowledgements}

The authors are grateful to the referee Alessandro Sonnenfeld for the thorough and helpful comments.
CEP thanks Leon Doddema, Martin Vogelaar and Ewout Helmich for help and support.
CEP, CT, GV, and LVEK are supported through an NWO-VICI grant (project number 639.043.308). CT also acknowledges funding from the INAF PRIN-SKA 2017 program 1.05.01.88.04. SC has been financially supported by a grant (project number 614.001.206) from the Netherlands Organization for Scientific Research (NWO).
GVK acknowledges financial support from the Netherlands Research School for Astronomy (NOVA) and Target. Target is supported by Samenwerkingsverband Noord Nederland, European fund for regional development, Dutch Ministry of economic affairs, Pieken in de Delta, Provinces of Groningen and Drenthe. NRN acknowledges financial support from the European Union Horizon 2020 research and innovation programme under the Marie Sklodowska-Curie grant agreement N. 721463 to the SUNDIAL ITN network.
This work is supported by the Deutsche Forschungsgemeinschaft in the framework of the TR33 `The Dark Universe'.
MB is supported by the Netherlands Organization for Scientific Research, NWO, through grant number 614.001.451 and by the Polish Ministry of Science and Higher Education through grant DIR/WK/2018/12.
BG acknowledges support from the European Research Council under grant number 647112.
JTAdJ is supported by the Netherlands Organisation for Scientific Research (NWO) through grant 621.016.402.
CH acknowledges support from the European Research Council under grant number 647112.
KK acknowledges support by the Alexander von Humboldt Foundation.
CS has received funding from the European Union's Horizon 2020 research and innovation programme under the Marie Sklodowska-Curie actions grant agreement No 664931.
This work is based on data products from observations made with ESO Telescopes at the La Silla Paranal Observatory under programme IDs 177.A-3016, 177.A-3017, and 177.A-3018, and on data products produced by Target/OmegaCEN, INAF-OACN, INAF-OAPD, and the KiDS production team, on behalf of the KiDS consortium. OmegaCEN and the KiDS production team acknowledge support by NOVA and NWO-M grants. Members of INAF-OAPD and INAF-OACN also acknowledge the support from the Department of Physics and Astronomy of the University of Padova, and of the Department of Physics of University of Federico II (Naples). 
GAMA is a joint European-Australasian project based around a spectroscopic campaign using the Anglo-Australian Telescope. The GAMA input catalogue is based on data taken from the SDSS and the UKIDSS. Complementary imaging of the GAMA regions is being obtained by a number of independent survey programmes including GALEX MIS, VST KiDS, VISTA VIKING, WISE, Herschel-ATLAS, GMRT, and ASKAP, providing UV to radio coverage. GAMA is funded by the STFC (UK), the ARC (Australia), the AAO, and the participating institutions. The GAMA website is www.gama-survey.org. 
Funding for the Sloan Digital Sky Survey IV has been provided by the Alfred P. Sloan Foundation, the U.S. Department of Energy Office of Science, and the Participating Institutions. SDSS acknowledges support and resources from the Center for High-Performance Computing at the University of Utah. The SDSS web site is www.sdss.org.
SDSS is managed by the Astrophysical Research Consortium for the Participating Institutions of the SDSS Collaboration including the Brazilian Participation Group, the Carnegie Institution for Science, Carnegie Mellon University, the Chilean Participation Group, the French Participation Group, Harvard-Smithsonian Center for Astrophysics, Instituto de Astrofisica de Canarias, The Johns Hopkins University, Kavli Institute for the Physics and Mathematics of the Universe (IPMU) / University of Tokyo, the Korean Participation Group, Lawrence Berkeley National Laboratory, Leibniz Institut für Astrophysik Potsdam (AIP), Max-Planck-Institut fur Astronomie (MPIA Heidelberg), Max-Planck-Institut fur Astrophysik (MPA Garching), Max-Planck-Institut fur Extraterrestrische Physik (MPE), National Astronomical Observatories of China, New Mexico State University, New York University, University of Notre Dame, Observatorio Nacional / MCTI, The Ohio State University, Pennsylvania State University, Shanghai Astronomical Observatory, United Kingdom Participation Group, Universidad Nacional Autonoma de Mexico, University of Arizona, University of Colorado Boulder, University of Oxford, University of Portsmouth, University of Utah, University of Virginia, University of Washington, University of Wisconsin, Vanderbilt University, and Yale University.
%This publication has been made possible by the participation in the Galaxy Zoo project of more than 20.000 volunteers from around the world, with almost 2 million classifications provided. Their contributions are individually acknowledged at http://authors.galaxyzoo.org/. The data are generated via the Zooniverse.org platform, development of which is funded by generous support, including a Global Impact Award from Google, and by a grant from the Alfred P. Sloan Foundation.
The \textsc{TOPCAT} \citep{TOPCAT} and \textsc{STILTS} \citep{STILTS} software have been extensively used in this project.
{\small \textit{Author Contributions:} All authors contributed to the development and writing of this paper. The authorship list is given in three groups: the lead authors (CEP,CT,GV,LVEK,GVK), followed by two alphabetical groups. The first alphabetical group includes those who are key contributors to both the scientific analysis and the data products. The second group covers those who have either made a significant contribution to the data products, or to the scientific analysis.}

%%%%%%%%%%%%%%%%%%%%%%%%%%%%%%%%%%%%%%%%%%%%%%%%%%



\bibliographystyle{mnras}
\bibliography{petrillo} % if your bibtex file is called example.bib

%%%%%%%%%%%%%%%%%%%%%%%%%%%%%%%%%%%%%%%%%%%%%%%%%%

%%%%%%%%%%%%%%%%% APPENDICES %%%%%%%%%%%%%%%%%%%%%
\appendix

\section{Data}\label{SECtoplinks}

\subsection{LinKS sample}
In the online material\footnote{\href{http://www.astro.rug.nl/lensesinkids}{\tt http://www.astro.rug.nl/lensesinkids }} we provide a table of our lens candidates properties, with:
%\footnote{\href{http://www.astro.rug.nl/lensesinkids}{\tt http://www.astro.rug.nl/lensesinkids }} 
%where for each of the 1983 candidates in the LinKS sample we report:

\begin{itemize}
\item an internal ID;
\item a score from the visual inspection;
\item the $p$-values from the ConvNets;
\item the lens-candidate coordinates;
\item a flag that indicates whether the candidate is already a confirmed lens or it has been identified as a candidate in other surveys.
\end{itemize}
In addition, at \href{http://www.astro.rug.nl/lensesinkids}{\tt http://www.astro.rug.nl/lensesinkids } we list for each one of the 1983 \textit{LinKS} candidates:
\begin{itemize}
\item the internal ID;
\item the visual inspection score; 
\item the lens-candidate coordinates;
\item the RGB stamp of $101 \times 101$ pixels, corresponding to $\sim20 \times 20$ arcseconds;   
\item a link to download their respective \textit{g}, \textit{r} an \textit{i} {\tt fits} files.
\end{itemize}
The candidates are ordered by decreasing visual inspection score.
We also present the RGB images of the 89 ``bona fide" candidates in \Fig\ref{FIGfirstcandidates}.

\captionsetup[subfigure]{labelformat=empty}
\begin{figure*}
\begin{center}
\subfloat[70]{\includegraphics[width=30mm]{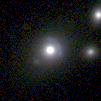}}
\subfloat[70]{\includegraphics[width=30mm]{1588.png}}
\subfloat[70]{\includegraphics[width=30mm]{2854.png}}
\subfloat[70]{\includegraphics[width=30mm]{2075.png}}
\subfloat[64]{\includegraphics[width=30mm]{51.png}}
\subfloat[64]{\includegraphics[width=30mm]{2251.png}}
\\
\subfloat[64]{\includegraphics[width=30mm]{3096.png}}
\subfloat[64]{\includegraphics[width=30mm]{2769.png}}
\subfloat[58]{\includegraphics[width=30mm]{1642.png}}
\subfloat[58]{\includegraphics[width=30mm]{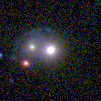}}
\subfloat[58]{\includegraphics[width=30mm]{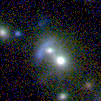}}
\subfloat[58]{\includegraphics[width=30mm]{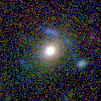}}
\\        
\subfloat[58]{\includegraphics[width=30mm]{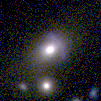}}
\subfloat[58]{\includegraphics[width=30mm]{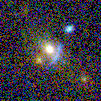}}
\subfloat[58]{\includegraphics[width=30mm]{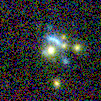}}
\subfloat[58]{\includegraphics[width=30mm]{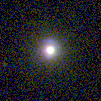}}
\subfloat[54]{\includegraphics[width=30mm]{2558.png}}
\subfloat[52]{\includegraphics[width=30mm]{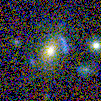}}
\\
\subfloat[52]{\includegraphics[width=30mm]{3154.png}}
\subfloat[48]{\includegraphics[width=30mm]{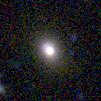}}
\subfloat[48]{\includegraphics[width=30mm]{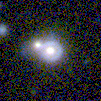}}
\subfloat[48]{\includegraphics[width=30mm]{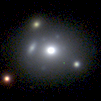}}
\subfloat[46]{\includegraphics[width=30mm]{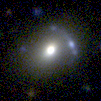}}
\subfloat[46]{\includegraphics[width=30mm]{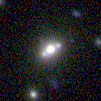}}
\\
\subfloat[44]{\includegraphics[width=30mm]{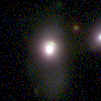}}
\subfloat[42]{\includegraphics[width=30mm]{1938.png}}
\subfloat[42]{\includegraphics[width=30mm]{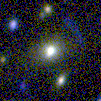}}
\subfloat[42]{\includegraphics[width=30mm]{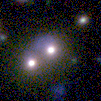}}
\subfloat[42]{\includegraphics[width=30mm]{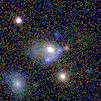}}
\subfloat[42]{\includegraphics[width=30mm]{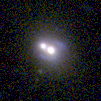}}
\\
\subfloat[42]{\includegraphics[width=30mm]{1231.png}}
\subfloat[40]{\includegraphics[width=30mm]{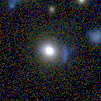}}
\subfloat[40]{\includegraphics[width=30mm]{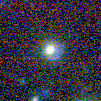}}
\subfloat[38]{\includegraphics[width=30mm]{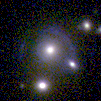}}
\subfloat[38]{\includegraphics[width=30mm]{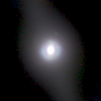}}
\subfloat[36]{\includegraphics[width=30mm]{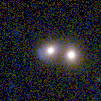}}
\caption{Images of the 89 candidates in the LinKS with a visual inspection score greater than 27. Each image has dimensions $20 \times 20$ arcseconds.}
\label{FIGfirstcandidates}
\end{center}
\end{figure*}

\captionsetup[subfigure]{labelformat=empty}
\begin{figure*}
\begin{center}
\subfloat[36]{\includegraphics[width=30mm]{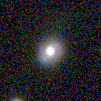}}
\subfloat[36]{\includegraphics[width=30mm]{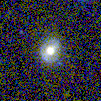}}
\subfloat[36]{\includegraphics[width=30mm]{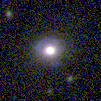}}
\subfloat[36]{\includegraphics[width=30mm]{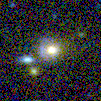}}
\subfloat[36]{\includegraphics[width=30mm]{2769.png}}
\subfloat[36]{\includegraphics[width=30mm]{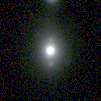}}
\\         
\subfloat[36]{\includegraphics[width=30mm]{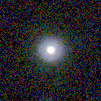}}
\subfloat[36]{\includegraphics[width=30mm]{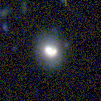}}
\subfloat[34]{\includegraphics[width=30mm]{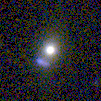}}
\subfloat[34]{\includegraphics[width=30mm]{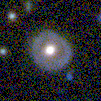}}
\subfloat[34]{\includegraphics[width=30mm]{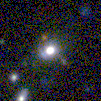}}
\subfloat[34]{\includegraphics[width=30mm]{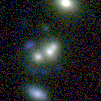}}
\\
\subfloat[32]{\includegraphics[width=30mm]{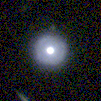}}
\subfloat[32]{\includegraphics[width=30mm]{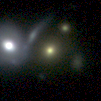}}
\subfloat[32]{\includegraphics[width=30mm]{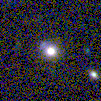}}
\subfloat[32]{\includegraphics[width=30mm]{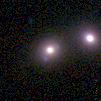}}
\subfloat[32]{\includegraphics[width=30mm]{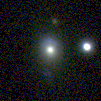}}
\subfloat[32]{\includegraphics[width=30mm]{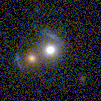}}
\\        
\subfloat[32]{\includegraphics[width=30mm]{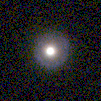}}
\subfloat[32]{\includegraphics[width=30mm]{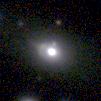}}
\subfloat[30]{\includegraphics[width=30mm]{844.png}}
\subfloat[30]{\includegraphics[width=30mm]{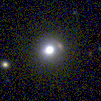}}
\subfloat[30]{\includegraphics[width=30mm]{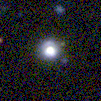}}
\subfloat[30]{\includegraphics[width=30mm]{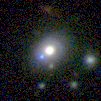}}
\\        
\subfloat[30]{\includegraphics[width=30mm]{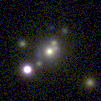}}
\subfloat[30]{\includegraphics[width=30mm]{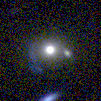}}
\subfloat[30]{\includegraphics[width=30mm]{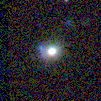}}
\subfloat[30]{\includegraphics[width=30mm]{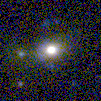}}
\subfloat[30]{\includegraphics[width=30mm]{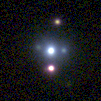}}
\subfloat[30]{\includegraphics[width=30mm]{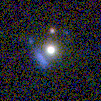}}
\\        
\subfloat[30]{\includegraphics[width=30mm]{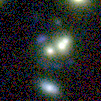}}
\subfloat[30]{\includegraphics[width=30mm]{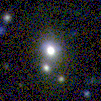}}
\subfloat[30]{\includegraphics[width=30mm]{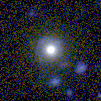}}
\subfloat[30]{\includegraphics[width=30mm]{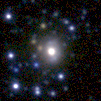}}
\subfloat[30]{\includegraphics[width=30mm]{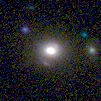}}
\subfloat[30]{\includegraphics[width=30mm]{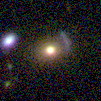}}
\hfill
\contcaption{}
\end{center}
\end{figure*}

\captionsetup[subfigure]{labelformat=empty}
\begin{figure*}
\begin{center}
\subfloat[30]{\includegraphics[width=30mm]{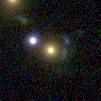}}
\subfloat[28]{\includegraphics[width=30mm]{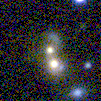}}
\subfloat[28]{\includegraphics[width=30mm]{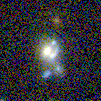}}
\subfloat[28]{\includegraphics[width=30mm]{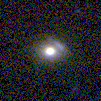}}
\subfloat[28]{\includegraphics[width=30mm]{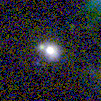}}
\subfloat[28]{\includegraphics[width=30mm]{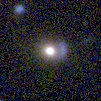}}
\\        
\subfloat[28]{\includegraphics[width=30mm]{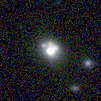}}
\subfloat[28]{\includegraphics[width=30mm]{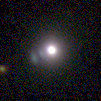}}
\subfloat[28]{\includegraphics[width=30mm]{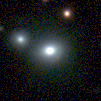}}
\subfloat[28]{\includegraphics[width=30mm]{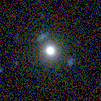}}
\subfloat[28]{\includegraphics[width=30mm]{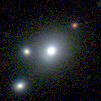}}
\subfloat[28]{\includegraphics[width=30mm]{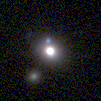}}
\\        
\subfloat[28]{\includegraphics[width=30mm]{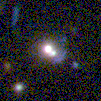}}
\subfloat[28]{\includegraphics[width=30mm]{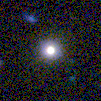}}
\subfloat[28]{\includegraphics[width=30mm]{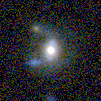}}
\subfloat[28]{\includegraphics[width=30mm]{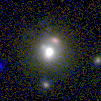}}
\subfloat[28]{\includegraphics[width=30mm]{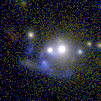}}
\hfill
\contcaption{}
\end{center}
\end{figure*}
%\onecolumn
%\begin{longtable}{lccccccccccr}
% \hline
% \input{table_candidates.tex}
% \hline
%\label{TABcandidates}
%\end{longtable}
%\clearpage
%\twocolumn

\subsection{Bonus sample}
%Here we present a collection of 200 candidates that are not part of the LRG sample and were retrieved serendipitously or, in great majority, during our experimentations with ConvNets. We name this collection of promising candidates ``Bonus sample".
The \textit{Bonus sample} is 
%presented in the online material and 
available via \href{http://www.astro.rug.nl/lensesinkids}{\tt http://www.astro.rug.nl/lensesinkids }
%\footnotemark[5]
similarly to the \textit{LinKS sample}.

%%%%%%%%%%%%%%%%%%%%%%%%%%%%%%%%%%%%%%%%%%%%%%%%%%

% Don't change these lines
\bsp	% typesetting comment
\label{lastpage}
\end{document}